# SCALING PROPERTIES OF THE ISING MODEL IN A FIELD

UWE GRIMM, BERNARD NIENHUIS

*Instituut voor Theoretische Fysica, Universiteit van Amsterdam,*
*Valckenierstraat 65, 1018 XE Amsterdam, The Netherlands*

The dilute $A_3$ model is a solvable IRF (interaction-round-a-face) model with three local states and adjacency conditions encoded by the Dynkin diagram of the Lie algebra $A_3$. It can be regarded as a solvable version of a critical Ising model in a magnetic field. One therefore expects the scaling limit to be governed by Zamolodchikov's integrable perturbation of the $c = 1/2$ conformal field theory. We perform a detailed numerical investigation of the solutions of the Bethe ansatz equation for the off-critical model. Our results agree perfectly with the predicted values for the lowest masses of the stable particles and support the assumptions on the nature of the Bethe ansatz solutions which enter crucially in a recent thermodynamic Bethe ansatz calculation of the factorized scattering matrix.

## 1 Introduction

Despite the fact that the two-dimensional Ising model in a magnetic field has defied an analytic solution to this day, relatively much is known about the properties of this model. In particular, a celebrated result of Zamolodchikov[1,2] shows that a perturbation of the $c = 1/2$ conformal field theory[3,4] (which corresponds to the critical Ising model) with the spin density operator of conformal dimensions $(\Delta, \bar{\Delta}) = (1/16, 1/16)$ leads to an integrable quantum field theory which contains eight massive particles with factorized scattering,[5-7] the particle masses and the $S$-matrix elements being related to the exceptional Lie algebra $E_8$.

In an appropriate scaling limit, the Ising model in a magnetic field should be described by this integrable field theory, and the predictions for the lowest mass ratios have been examined numerically by several authors.[8-12] As these results rely on relatively small size transfer matrix calculations or on Monte Carlo simulations, they provide rather crude checks for the lowest mass ratios only.

The recent construction [13,14] of a solvable version of the Ising model in a field, the so-called dilute $A_3$ model, gives us the opportunity to have a closer look at this scaling limit. By a thermodynamic Bethe ansatz calculation,[15] it has indeed been possible to recover Zamolodchikov's result for the mass ratios and the $S$-matrix. This calculation relies on assumptions concerning the string structure of the solutions of the Bethe ansatz equations for the dilute $A_3$ model. However, a first numerical check shows that the largest eigenvalues of the transfer matrix at criticality do not conform with the predicted string structures. This observation has motivated us to perform an extensive numerical investigation of the corresponding off-critical Bethe ansatz equations, to obtain precise estimates of some of the mass ratios on the one hand and in order to check the assumptions on the structure of its solutions on the other.

This article is organized as follows. We commence by introducing the dilute $A_3$ model. In Sec. 3, the Bethe ansatz equations for the dilute $A_3$ model are given. Subsequently, we discuss the critical behaviour of the model, and in Sec. 5 the scaling limit is defined. In Sec. 6, we present our numerical results and we conclude in Sec. 7.

## 2  The dilute $A_3$ model

The so-called dilute A–D–E models are a series of solvable lattice models (in the sense of commuting transfer matrices) which have been constructed recently.[16,13] They are critical IRF (interaction-round-a-face) models with local states and adjacency conditions encoded by Dynkin diagrams of the simply laced Lie algebras. The dilute $A_L$ models can be extended[13] to off-critical solutions of the Yang-Baxter equation with Boltzmann weights parametrized in terms of elliptic functions of the spectral parameter. These models are of particular interest for an odd number of local states $L$. In contrast to the critical case, the off-critical weights of these models do not respect the $\mathbb{Z}_2$-symmetry of the underlying Dynkin diagram. This means that the solution describes the critical system in the presence of a symmetry-breaking field rather than a temperature-like deviation from criticality.

Here, we are only interested in the dilute $A_3$ model, which is a three-state model built on the effective adjacency diagram depicted in Fig. 1. In an allowed configuration on the square lattice, all pairs of states on neighbouring lattice sites have to be directly connected on the adjacency diagram. In other words, the states '+' and '−' cannot occur on adjacent lattice sites in any allowed configuration. The model is defined by associating Boltzmann weights

$$W \left( \begin{array}{cc} d & c \\ a & b \end{array} \middle| u \right) = \begin{array}{c} d \quad\quad\quad c \\ \boxed{\phantom{xx} u \phantom{xx}} \\ a \quad\quad\quad b \end{array} \quad\quad (1)$$

to each configuration of states $a, b, c, d$ around the elementary plaquettes of the square lattice. As the explicit form of the weights is rather cumbersome,[13] we refrain from quoting these here and follow the notation of Refs. 15, 17 and 18. Let us only mention that besides their dependence on the spectral parameter $u$, the weights depend on the elliptic nome $q$ (with $q = 0$ corresponding to the critical case). For fixed $q$, they satisfy the Yang-Baxter equation

$$\sum_g \begin{array}{c}\text{(LHS diagram)}\end{array} = \sum_g \begin{array}{c}\text{(RHS diagram)}\end{array} \qquad (2)$$

which implies [19] that the row transfer matrices $T(u)$ form a one-parameter family of commuting matrices, i.e., $[T(u), T(v)] = 0$.

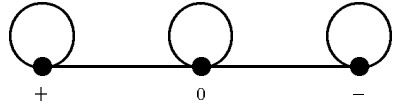

Figure 1: Effective adjacency diagram of the dilute $A_3$ model.

## 3  Bethe Ansatz Solution

The dilute $A_L$ models can be solved by a Bethe ansatz. [15,17] The eigenvalues $\Lambda(u)$ of the row transfer matrix $T(u)$ (with periodic boundary conditions) are given by

$$\begin{aligned}\Lambda(u) = &\; \omega \left(-\frac{\vartheta_1(u-2\lambda)\,\vartheta_1(u-3\lambda)}{\vartheta_1(2\lambda)\,\vartheta_1(3\lambda)}\right)^N \prod_{j=1}^N \frac{\vartheta_1(u-u_j+\lambda)}{\vartheta_1(u-u_j-\lambda)} \\ &+ \left(-\frac{\vartheta_1(u)\,\vartheta_1(u-3\lambda)}{\vartheta_1(2\lambda)\,\vartheta_1(3\lambda)}\right)^N \prod_{j=1}^N \frac{\vartheta_1(u-u_j)\,\vartheta_1(u-u_j-3\lambda)}{\vartheta_1(u-u_j-\lambda)\,\vartheta_1(u-u_j-2\lambda)} \\ &+ \omega^{-1} \left(-\frac{\vartheta_1(u)\,\vartheta_1(u-\lambda)}{\vartheta_1(2\lambda)\,\vartheta_1(3\lambda)}\right)^N \prod_{j=1}^N \frac{\vartheta_1(u-u_j-4\lambda)}{\vartheta_1(u-u_j-2\lambda)} \end{aligned} \qquad (3)$$

where $N$ denotes the number of faces in the row, and

$$\lambda = 5\pi/16, \qquad \omega = \exp(i\pi\ell/4), \qquad 0 < u < 3\lambda, \qquad (4)$$

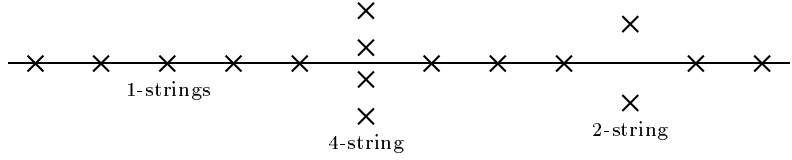

Figure 2: Sketch of typical arrangements of Bethe ansatz roots $v_j$ in the complex plane.

where $\ell \in \{1, 2, 3\}$. The Bethe ansatz roots $u_j$ are determined by the Bethe ansatz equations

$$\omega^\ell \left( \frac{\vartheta_1(u_j - \lambda)}{\vartheta_1(u_j + \lambda)} \right)^N = - \prod_{k=1}^{N} \frac{\vartheta_1(u_j - u_k - 2\lambda)\, \vartheta_1(u_j - u_k + \lambda)}{\vartheta_1(u_j - u_k + 2\lambda)\, \vartheta_1(u_j - u_k - \lambda)} \quad (5)$$

with $j = 1, 2, \ldots, N$.

Defining $v_j = i u_j$, the largest eigenvalue $\Lambda_0(u)$ (in the sector $\ell = 1$) corresponds to a set of $N$ real Bethe ansatz roots $v_j$. In general, the solutions will however be complex, but for large system size $N$ they typically arrange into so-called *strings* (subsets of roots with approximately the same real part) as shown schematically in Fig. 2, where the horizontal line represents the real axis. It should be noted that the 'string type' of a particular solution might well depend on the variable $q$, an example of this behaviour will be given below.

## 4 Critical Behaviour

As already mentioned, the dilute $A_3$ model is critical for $q = 0$. In the regime under consideration, see Eq. 4, the critical point is of Ising type and the model is therefore described by a $c = 1/2$ conformal field theory with scalar primary fields of conformal dimensions $(\Delta, \bar{\Delta})$ with $\Delta = \bar{\Delta} = 0$ (corresponding to the identity operator), $\Delta = \bar{\Delta} = 1/16$ (the spin density operator) and $\Delta = \bar{\Delta} = 1/2$ (the energy density operator). Each of these primary fields gives rise to an infinite tower of descendant fields with conformal dimensions $(\Delta + r, \bar{\Delta} + \bar{r})$, where $r$ and $\bar{r}$ are positive integers, which together form the product of the corresponding irreducible representations of the Virasoro algebra.

As a consequence of conformal invariance, the appropriately scaled spectral gaps

$$x_j = \frac{N}{2\pi} \log(\Lambda_0^{(0)} / \Lambda_j^{(0)}) \quad (6)$$

in the critical limit ($q = 0$, $N \to \infty$) have the form $\Delta + r + \bar{\Delta} + \bar{r}$ with the aforementioned values. Here, we denoted by $\Lambda_j^{(0)}$ the eigenvalues of the transfer matrix $T(3\lambda/2)$ at the isotropic point $u = 3\lambda/2$ for $q = 0$, $\Lambda_0^{(0)}$ being the largest eigenvalue. Moreover, the central charge $c = 1/2$ manifests itself in the finite-size corrections of the largest eigenvalue

$$-\log(\Lambda_0^{(0)}) = Nf_0 + \frac{\pi c}{6N} + o(N^{-1}). \tag{7}$$

Table 1: Finite-size approximants for central charge $c$ and smallest scaling dimensions $x_j$.

| $N$ | $c$ | $x_1$ | $x_2$ | $x_3$ | $x_4$ | $x_5$ | $x_6$ | $x_7$ |
|---|---|---|---|---|---|---|---|---|
| 10 | 0.499681 | 0.1250802 | 1.0145 | 2.196 | 3.148 | 4.272 | 4.417 | 5.449 |
| 20 | 0.499920 | 0.1250200 | 1.0035 | 2.142 | 3.034 | 4.062 | 4.191 | 5.098 |
| 50 | 0.499987 | 0.1250032 | 1.0006 | 2.128 | 3.005 | 4.010 | 4.135 | 5.015 |
| 75 | 0.499994 | 0.1250014 | 1.0003 | 2.126 | 3.002 | 4.004 | 4.130 | 5.007 |
| 100 | 0.499997 | 0.1250008 | 1.0001 | 2.126 | 3.001 | 4.002 | 4.128 | 5.004 |
| $\infty$ | 1/2 | 1/8 | 1 | $2\,1/8$ | 3 | 4 | $4\,1/8$ | 5 |

In Table 1, numerical approximants for the smallest scaling dimensions $x_j$ of zero momentum operators (i.e., $\Delta + r = \bar{\Delta} + \bar{r}$) are presented for several system sizes $N$. These are obtained from Eq. 6, where the eigenvalues $\Lambda_j^{(0)}$ have been calculated by solving the Bethe ansatz equations numerically, see Sec. 6 for details. Furthermore, Table 1 also includes finite-size approximants for the central charge $c = 1/2$ obtained by neglecting the $o(N^{-1})$ correction terms in Eq. 7, where the known analytic expression [18] for the specific free energy $f_0$ has been used.

## 5 Scaling Limit

The additional parameter $q$ (which has the physical interpretation of a magnetic field) allows us to approach the critical point in different ways, resulting in a *scaling limit* characterized by the *scaling variable* $\mu$

$$\mu = q\,N^{15/8}, \tag{8}$$

i.e., the limits $q \to 0$ and $N \to \infty$ are taken simultaneously keeping $\mu$ constant. Here, $\mu = 0$ corresponds to the critical limit discussed above, $\mu = \infty$ to the massive limit where Zamolodchikov's $E_8$ field theory applies, see Fig. 3. The appropriately scaled spectral gaps

$$f_j = q^{-8/15}\,\log(\Lambda_0/\Lambda_j) \tag{9}$$

in the scaling limit become functions of the scaling variable $\mu$ alone.

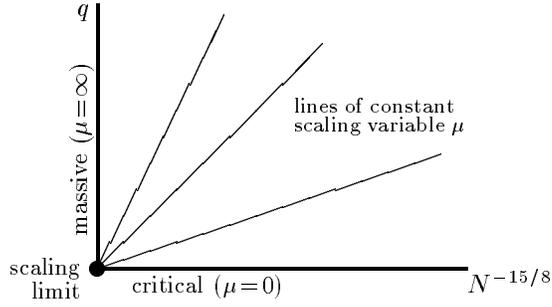

Figure 3: Schematic view of the scaling limit in the variables $q$ and $N$.

In the massive limit, the ratios $r_j = f_{j+1}/f_1$ supposedly approach the particle mass ratios of the corresponding massive field theory. The mass ratios of the eight stable particles are determined by the entries of the Perron-Frobenius eigenvector of the Cartan matrix of the Lie algebra $E_8$ and (ordered by magnitude) are given by

$$\begin{aligned}
m_1/m &= 1 \\
m_2/m &= 2\cos(\pi/5) &= 1.618034\ldots \\
m_3/m &= 2\cos(\pi/30) &= 1.989044\ldots \\
m_4/m &= 4\cos(\pi/5)\cos(7\pi/30) &= 2.404867\ldots \\
m_5/m &= 4\cos(\pi/5)\cos(2\pi/15) &= 2.956295\ldots \\
m_6/m &= 4\cos(\pi/5)\cos(\pi/30) &= 3.218340\ldots \\
m_7/m &= 8\cos^2(\pi/5)\cos(7\pi/30) &= 3.891157\ldots \\
m_8/m &= 8\cos^2(\pi/5)\cos(2\pi/15) &= 4.783386\ldots
\end{aligned} \quad (10)$$

where the corresponding enumeration of nodes in the Dynkin diagram of $E_8$ is shown in Fig. 4.

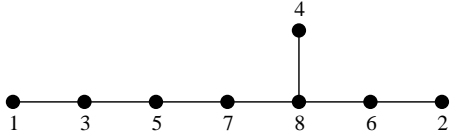

Figure 4: Dynkin diagram of $E_8$ with nodes labelled according to the masses of Eq. 10.

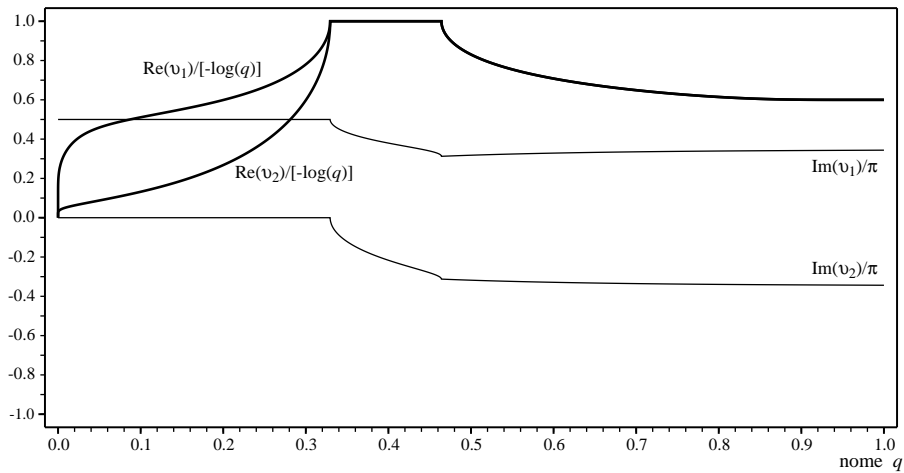

Figure 5: Bethe ansatz roots for the second largest eigenvalue for system size $N = 2$. The scales are normalized to the periods for both the real and the imaginary part.

## 6 Numerical Results

For the numerical treatment of the Bethe ansatz equations we employed a modified Newton method[20]. We took the following course of action: As a first step, the eigenvalues of the transfer matrix were obtained by direct diagonalization for small system sizes (up to $N = 10$). Secondly, the corresponding solutions of the Bethe ansatz equations were identified and generalized to larger $N$. Finally, these solutions were then followed as the variable $q$ is varied.

Since we are mainly interested in the masses of single particle states, we restricted ourselves to the zero momentum sector, i.e. to translation invariant eigenvectors of the transfer matrix. The numerical data of Table 1 confirm that we correctly identified the states with scaling dimensions $x \leq 5$ in the critical limit. Apart from the largest eigenvalue, which corresponds to a set of $N$ real Bethe ansatz roots $v_j$ for all values of $q$ ($0 \leq q < 1$), the root arrangements of all the other eigenvalues change several times as $q$ is varied. A simple case is shown in Fig. 6, where the Bethe ansatz roots for the second largest eigenvalue (sector $\ell = 2$) are given for $N = 2$. Comparing different system sizes, one finds that these changes actually occur at approximatively constant values of the scaling parameter $\mu$, so they do enter in the scaling limit. Of course, the corresponding finite-size eigenvalues are analytic functions of $q$ always.

As mentioned above, the thermodynamic Bethe ansatz calculation[15] relies on assumptions on the string structure of the Bethe ansatz solutions. For the eigenvalues under consideration, we find that for sufficiently large scaling variable $\mu$ these assumptions are indeed correct, except for one eigenvalue (corresponding to $\Delta = \bar{\Delta} = 1/2$ and $r = \bar{r} = 1$ ($x = 3$), respectively to the mass $m_4$) where one might have to go to much larger values of the scaling variable than we did. Therefore, the apparent conflict mentioned in Sec. 1, between the string types which are observed at criticality and those which enter the thermodynamic Bethe ansatz approach, appears to be solved.

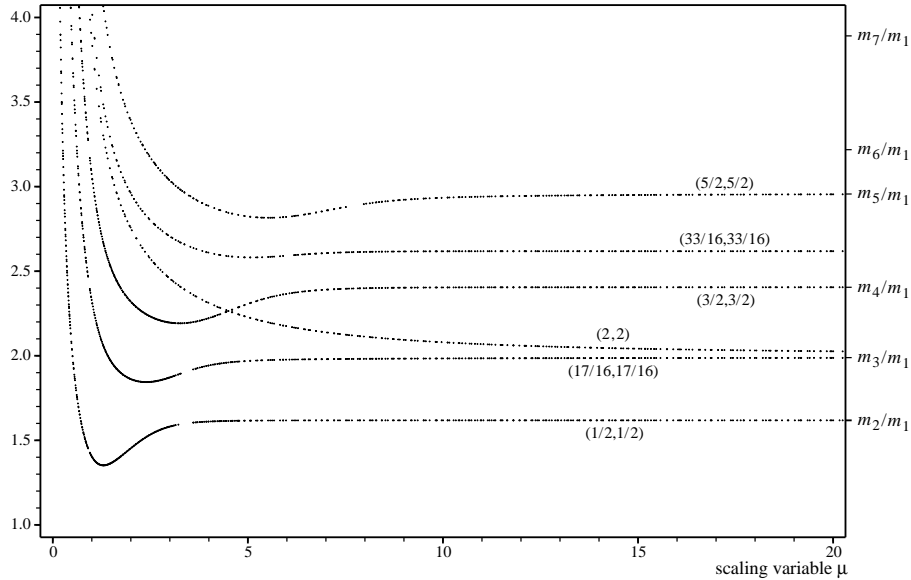

Figure 6: Ratios of scaled gaps as functions of the scaling parameter $\mu$. The corresponding scaling dimensions $(\Delta + r, \bar{\Delta} + \bar{r})$ are indicated.

In Fig. 6, the dependence of the ratios $r_j = f_{j+1}/f_1$ of the scaled gaps $f_j$ (Eq. 9) on the scaling variable $\mu$ is shown. In this plot, we used data from system sizes $N = 50$, $N = 75$ and $N = 100$, they clearly scale as expected. This is also supported by the numerical data for three values of scaling parameters $\mu$ given in Table 2. The convergence to the limit values is quite obvious, and one recognizes four non-trivial mass ratios as well as two ratios belonging to two-particle states (with masses $2m_1$ and $m_1 + m_2$, respectively). Also, among the scaling functions considered we observe one intersection (between the functions labelled by conformal dimensions $(3/2, 3/2)$ (mass $m_4$) and $(2, 2)$ (mass $2m_1$), respectively), in agreement with results obtained by a truncated fermionic space approach.[21] Clearly, our results also agree with the previous numerical investigations of the Ising model in a magnetic field,[8-12] which however are much less accurate. One disagreement is due to an error: Given the order of masses in equation (1.10) of Sagdeev and Zamolodchikov,[9] the critical limit (their equation (1.9)) should read $\Delta_0 = 0$, $\Delta_1 = 1/16$, $\Delta_2 = 1/2$, $\Delta_3 = 17/16$, $\Delta_4 = 2$.

Table 2: Ratios of scaled gaps for three values of the scaling parameter $\mu$.

| $\mu$ | $N$ | $r_1$ | $r_2$ | $r_3$ | $r_4$ | $r_5$ | $r_6$ |
|---|---|---|---|---|---|---|---|
| 10 | 10 | 1.618749 | 1.983943 | 2.403558 | 2.083461 | 2.616910 | 2.910107 |
|  | 20 | 1.618037 | 1.983923 | 2.403649 | 2.081017 | 2.617335 | 2.930942 |
|  | 50 | 1.618024 | 1.983977 | 2.403797 | 2.079804 | 2.617468 | 2.934096 |
|  | 75 | 1.618024 | 1.983984 | 2.403815 | 2.079550 | 2.617482 | 2.934583 |
|  | 100 | 1.618025 | 1.983990 | 2.403834 | 2.079551 | 2.617497 | 2.934617 |
| 20 | 10 | 1.623692 | 1.989446 | 2.412441 | 2.017536 | 2.623691 | 2.958322 |
|  | 20 | 1.618144 | 1.987688 | 2.405009 | 2.025316 | 2.618144 | 2.954218 |
|  | 50 | 1.618034 | 1.987597 | 2.404866 | 2.026212 | 2.618034 | 2.954289 |
|  | 75 | 1.618034 | 1.987591 | 2.404865 | 2.026286 | 2.618034 | 2.954306 |
|  | 100 | 1.618034 | 1.987589 | 2.404865 | 2.026315 | 2.618034 | 2.954311 |
| 30 | 10 | 1.634823 | 1.992979 | 2.428402 | 2.003678 | 2.634823 | 2.970263 |
|  | 20 | 1.618424 | 1.988671 | 2.405376 | 2.012146 | 2.618424 | 2.956329 |
|  | 50 | 1.618036 | 1.988497 | 2.404869 | 2.013483 | 2.618036 | 2.956009 |
|  | 75 | 1.618034 | 1.988489 | 2.404867 | 2.013604 | 2.618034 | 2.956007 |
|  | 100 | 1.618034 | 1.988486 | 2.404867 | 2.013644 | 2.618034 | 2.956007 |
| $\infty$ | $\infty$ | 1.618034 | 1.989044 | 2.404867 | 2.000000 | 2.618034 | 2.956295 |

## 7 Concluding Remarks

The Bethe ansatz solutions of the off-critical dilute $A_3$ model were studied numerically. Our results support the equivalence to Zamolodchikov's magnetic Ising model [1,2] as well as recent assumptions on the string type of the solutions which enter a thermodynamic Bethe ansatz calculation [15] of the $S$-matrix. A more detailed account of our results will be published elsewhere.[22]

**Acknowledgments**

This work has been supported by Samenwerkingsverband FOM/SMC Mathematische Fysica.